\documentclass[twocolumn,aps,prl,superscriptaddress,showpacs,floatfix,longbibliography]{revtex4-1}
\usepackage[colorlinks,linkcolor=blue,citecolor=blue,filecolor=black,urlcolor=blue]{hyperref}
\usepackage{epsfig,graphics}
\usepackage{graphicx}
\usepackage{dcolumn}
\usepackage{bm}
\usepackage{amssymb}
\usepackage{amsmath}
\usepackage{multirow}
\usepackage{float}
\usepackage{harpoon}
\usepackage{MnSymbol}
\usepackage{appendix}
\usepackage{color}

\newcommand{\nch} {N_{\mathrm{ch}}}

\newcommand{\lr}[1]{\left\langle #1\right\rangle}

\newcommand{\npart}{N_{\mathrm{part}}}

\begin{document}
\title{Impact of the radial profile of atomic nuclei on observables in high-energy collisions}
\newcommand{\bnl}{Physics Department, Brookhaven National Laboratory, Upton, NY 11976, USA}
\newcommand{\sbu}{Department of Chemistry, Stony Brook University, Stony Brook, NY 11794, USA}
\newcommand{\tongji}{School of Physics Science and Engineering, Tongji University, Shanghai 200092, China}
\author{Zhengxi Yan}\email{zhengxi1yan@gmail.com}\affiliation{\sbu}
\author{Jun Xu}\email{junxu@tongji.edu.cn}\affiliation{\tongji}
\author{Jiangyong Jia}\email{jiangyong.jia@stonybrook.edu}\affiliation{\sbu}\affiliation{\bnl}\
\begin{abstract}
In heavy-ion phenomenology, the nucleon density distribution in colliding nuclei is commonly described by a two-parameter Woods-Saxon (WS) distribution. However, this approach overlooks the detailed radial structure in the density distribution that arises from the quantal filling patterns of neutrons and protons. These fine structures, as estimated by the Skyrme-Hartree-Fock density functional, cause slight deviations in heavy-ion observables from the WS baseline, which cannot be captured by simply readjusting the WS parameters. These deviations depend on centrality and observable but often exhibit similar shapes for different nuclei. To fully exploit the exceptional sensitivity of isobar collisions to nuclear structure, such fine structures should be considered.
\end{abstract}
\maketitle
High-energy nuclear collisions at RHIC and LHC create a hot and dense quark-gluon plasma (QGP), whose space-time expansion is described by relativistic viscous hydrodynamic equations~\cite{Heinz:2013th}. Due to the extremely short timescales of these collisions, they effectively capture a snapshot image of the nucleon spatial distribution in the nuclei, which generates the initial condition of the QGP. The initial condition, through hydrodynamic expansion, then impacts the particle momentum distributions measured in detectors~\cite{Ollitrault:2023wjk,STAR:2024eky}.

This imaging idea is particularly effective when comparing nuclei with similar masses but different internal structures~\cite{Giacalone:2021uhj}. Nuclear structure influences the geometric features of the initial condition, such as ellipticity $\varepsilon_2$, triangularity $\varepsilon_3$, and the number of participating nucleons $\npart$, which are linearly related to observables describing collective behavior, like elliptic flow $v_2$, triangular flow $v_3$, and charged particle multiplicity $\nch$\cite{Bozek:2012fw,Niemi:2012aj}. By constructing ratios of these observables between two systems, the effects of hydrodynamic response cancel, isolating the global shape and size features of the colliding nuclei~\cite{Zhang:2022fou}. Moreover, owing to the cancellation of experimental uncertainties, these ratios can be measured with great precision, percent-level or better~\cite{STAR:2021mii}, making them excellent probes for nuclear structure and initial condition~\cite{Bally:2022vgo}.

In high-energy phenomenology, the global density distribution of nuclei is often parameterized by a Woods-Saxon (WS) function, 
\begin{align}\label{eq:1}
\rho (r,\theta,\phi)) \propto [1+\exp(r-R_0(1+f(\theta,\phi))/a)]^{-1},
\end{align}
where $R_0$ and $a$ represent the half-density radius and skin thickness, respectively. The function $f(\theta,\phi)$ accounts for nuclear deformations. For example, for axial-symmetric quadrupole and octupole deformations, $f(\theta,\phi) = \beta_2 Y_{2,0}(\theta,\phi)+\beta_3 Y_{3,0}(\theta,\phi)$. The ratios of observables $\mathcal{O}=v_n$ and $\nch$ or $\varepsilon_n$ and $\npart$ between isobar systems X+X and Y+Y follow a scaling relation as a function of centrality~\cite{Jia:2021oyt}:
\begin{align}\label{eq:2}
D_{\mathcal{O}}\equiv\frac{\mathcal{O}_{\mathrm{\mathrm{X}}}}{\mathcal{O}_{\mathrm{\mathrm{Y}}}} -1 \approx c_1\Delta R_0 +c_2\Delta a + c_3 \Delta\beta_2^2 +c_4 \Delta\beta_3^2 \;,
\end{align}
where $\Delta R_0 =R_{0,\mathrm{\mathrm{X}}}-R_{0,\mathrm{\mathrm{Y}}}$, $\Delta a =a_{\mathrm{\mathrm{X}}}-a_{\mathrm{\mathrm{Y}}}$, and $\Delta\beta_n^2= \beta_{n,\mathrm{\mathrm{X}}}^2-\beta_{n,\mathrm{\mathrm{Y}}}^2$.

Recent STAR measurements in isobar and isobar-like systems~\cite{STAR:2021mii,STAR:2024eky} align well with this scaling. For example, in $^{96}$Ru+$^{96}$Ru and $^{96}$Zr+$^{96}$Zr collisions, the measured values of $D_{\mathcal{O}}$ range from 2--10\% depending on centrality with uncertainties less than 0.5\%, leading to a very accurate determination of WS parameters by comparing the $D_{\mathcal{O}}$ with hydrodynamic and transport model calculations~\cite{Jia:2021oyt,Nijs:2021kvn,STAR:2024eky}. The extracted values, $\Delta a = a_{\mathrm{Ru}}-a_{\mathrm{Zr}}= -0.06$~fm and $\Delta R_0=R_{0,\mathrm{Ru}}-R_{0,\mathrm{Zr}}=0.07$~fm, are consistent with the low-energy knowledge. 

However, there is a caveat. The realistic distributions of the protons and neutrons in the nucleus are more accurately described by Density Function Theories (DFT) based on effective nuclear forces~\cite{Bender:2003jk}. WS parameters derived from fitting DFT distributions miss detailed features~\cite{Xu:2021vpn}, particularly the wiggly structures associated with the orbital filling pattern of protons and neutrons. In the presence of nuclear deformation, the skin thickness may vary with the polar angle $\theta$~\cite{Liu:2023qeq}. As these fine features impact isobar ratios, not considering them will lead to uncertainties in extracting WS parameters via Eq.~\eqref{eq:2}.
 
This Letter aims to quantify the impact of the differences between DFT and WS on heavy-ion observables, and to compare these effects to the expected changes between isobar systems. We consider spherical nuclei and three observables: $\varepsilon_2$, $\varepsilon_3$, and $\npart$, for which Eq.~\eqref{eq:2} simplifies to:
\begin{align}\label{eq:3}
D_{\mathcal{O}}\approx  c_1\Delta R_0 +c_2\Delta a\;.
\end{align}
We consider several representative nuclei and assume that the structure differences from their isobar partners are similar to those between $^{96}$Ru and $^{96}$Zr. We found that the impact of DFT and WS differences on observables are usually smaller than those from assumed isobar differences of $\Delta a= -0.06$~fm and $\Delta R_0=0.07$~fm, but are not negligible for accurate extraction of nuclear structures with the above ratios.

{\bf Setup}. There are many DFTs with different forces one could choose from~\cite{Dutra:2012mb,Dutra:2014qga}. For this study, we use the standard Skyrme-Hartree-Fock (SHF) model~\cite{Chen:2010qx}. The SHF model defines the parameters of the Skyrme interaction in terms of ten macroscopic quantities, which have empirical values corresponding to the MSL0 force as in Ref.~\cite{Chen:2010qx,Liu:2022kvz}. We generate spatial distributions of protons and neutrons, summing them to obtain the radial distributions of nucleons $\rho(r)$. The parameters $R_0$ and $a$ are determined by matching the first and second radial moments of the WS distribution to those of the SHF~\cite{Xu:2021vpn}. The $n^{\mathrm{th}}$ moment is defined as,
\begin{align}\label{eq:4}
\left\langle r^n\right\rangle = \frac{\int r^{2+n}\rho(r) d r}{\int r^2 \rho(r) dr}\;.
\end{align}

\begin{figure}[!h]
\includegraphics[width=0.8\linewidth]{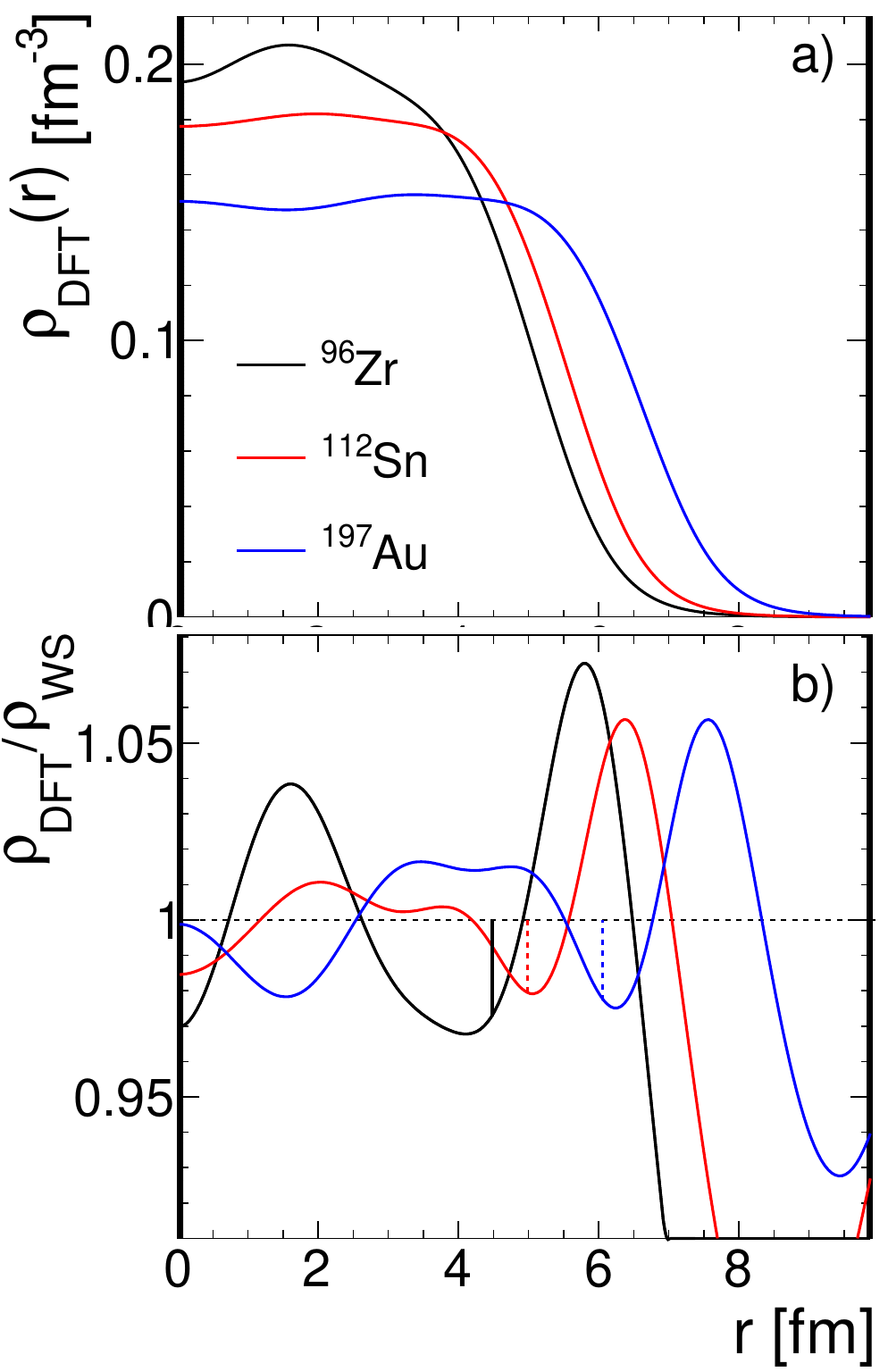}
\caption{\label{fig:1} (a) SHF nucleon density distributions for $^{96}$Zr, $^{112}$Sn, and $^{197}$Au nuclei. (b) Ratio of SHF and WS nucleon density distribution via moment matching. The vertical dashed lines indicate location $r=R_0-a$, where we combine the SHF to the left and WS to the right together to form $\rho_{\mathrm{mix}}$ (see text).}
\end{figure}

We scanned many near-spherical nuclei and compared their SHF distributions with the WS fit. From this comparison, we identified three nuclei with characteristically different shapes in $\rho(r)$: $^{96}$Zr, $^{112}$Sn and $^{197}$Au, as shown in Fig.~\ref{fig:1}. The $^{96}$Zr has a wiggle with a dip around $r=0$, while $^{197}$Au exhibits a wiggle with a peak around $r=0$, both showing significant deviations from the WS distributions. In contrast, $^{112}$Sn has a relatively uniform distribution with a smaller difference from the WS fit. The corresponding WS parameters are listed in Table~\ref{tab:1}.
\begin{table}[!h]
\begin{tabular}{ c|c|c|c}
          & $^{96}$Zr & $^{112}$Sn & $^{197}$Au\\\hline
$R_0$ (fm)& 5.059    &  5.538     & 6.610\\
$a$ (fm)  &  0.5774  & 0.5576     & 0.5552
\end{tabular}
\caption{Woods-Saxon parameters obtained by matching the first and second moments of nuclear distributions to the SHF.}
\label{tab:1}
\end{table}

We used a standard Monte-Carlo Glauber model~\cite{Miller:2007ri} to simulate collisions of $^{96}$Zr, $^{112}$Sn, and $^{197}$Au, assuming a nucleon-nucleon cross-section of 42 mb. Nucleons are modeled with a hard core, maintaining a minimum separation of 0.4 fm. For each species, 100 million minimum bias events were generated, separately for SHF and WS distributions, with their 1st and 2nd radial moments matched. These density distributions are labelled as $\rho_{\mathrm{shf}}$ and $\rho_{\mathrm{ws}}$, respectively. To isolate the impact of the wiggle structure, we constructed another $\rho(r)$ distribution by combining SHF at low radii ($r<R_0-a$) and WS at higher radii ($r>R_0-a$), the combined distribution is then rescaled to the mass number, labeled as $\rho_{\mathrm{mix}}$. The isobar partners for each species are mocked up by adjusting $a$ by -0.06 fm or $R_0$ by +0.07 fm, labeled as $\rho_{\mathrm{ws1}}$ and $\rho_{\mathrm{ws2}}$, respectively. Thus, five separate simulations were performed for each collision species as shown in Tab~\ref{tab:2}.  The difference between DFT and WS is accounted for by comparing $\rho_{\mathrm{shf}}$ or $\rho_{\mathrm{mix}}$ with $\rho_{\mathrm{ws}}$. The typical signal expected between a species and its isobar partner is accounted for by comparing $\rho_{\mathrm{ws1}}$ or $\rho_{\mathrm{ws2}}$ with $\rho_{\mathrm{ws}}$.

\begin{table}[!h]
\begin{tabular}{ c|c}
default       & $\rho_{\mathrm{ws}}$\\\hline
impact of DFT & $\rho_{\mathrm{shf}}$, $\rho_{\mathrm{mix}}$\\\hline
impact of isobars & $\rho_{\mathrm{ws1}}$, $\rho_{\mathrm{ws2}}$\\\hline
\end{tabular}
\caption{Five nucleon density distributions for studying the impact of wiggle structures from DFT and the typical difference between the isobar pairs.}
\label{tab:2}
\end{table}

For each event, we calculated $\npart$, $\varepsilon_2$, and $\varepsilon_3$, defined from the transverse nucleon density $\rho_{\perp}(r_\perp,\phi)=\int \rho dz$ as:
\small\begin{align}\label{eq:4b}
\varepsilon_n e^{in\Phi_n}\equiv -\frac{\int  \rho_{\perp}(r_{\perp},\phi)  r_{\perp}^n e^{in\phi} r_{\perp}dr_{\perp}d\phi}{\int  \rho_{\perp}(r_{\perp},\phi) r_{\perp}^n r_{\perp}dr_{\perp}d\phi}\;.
\end{align}\normalsize
We then constructed histograms of $\varepsilon_n \equiv \sqrt{\lr{\varepsilon_n^2}}$ and $p(\npart)$ as functions of $\npart$.  Note that the wiggly structures in $\rho_{\perp}(r_\perp)$ have smaller magnitudes than $\rho(r)$ in the inner region (Appendix). The distributions of each observable are divided by those from the default WS parameterization, resulting in four relative variations: $D_{\mathcal{O}}\{\mathrm{shf}\}$ and $D_{\mathcal{O}}\{\mathrm{mix}\}$ for evaluating the difference between DFT and WS, as well as $D_{\mathcal{O}}\{\mathrm{ws1}\}$ and $D_{\mathcal{O}}\{\mathrm{ws2}\}$ to account for the signal in isobar ratios. They are presented as a function of $\npart$.

\begin{figure*}[htbp]
\includegraphics[width=0.8\linewidth]{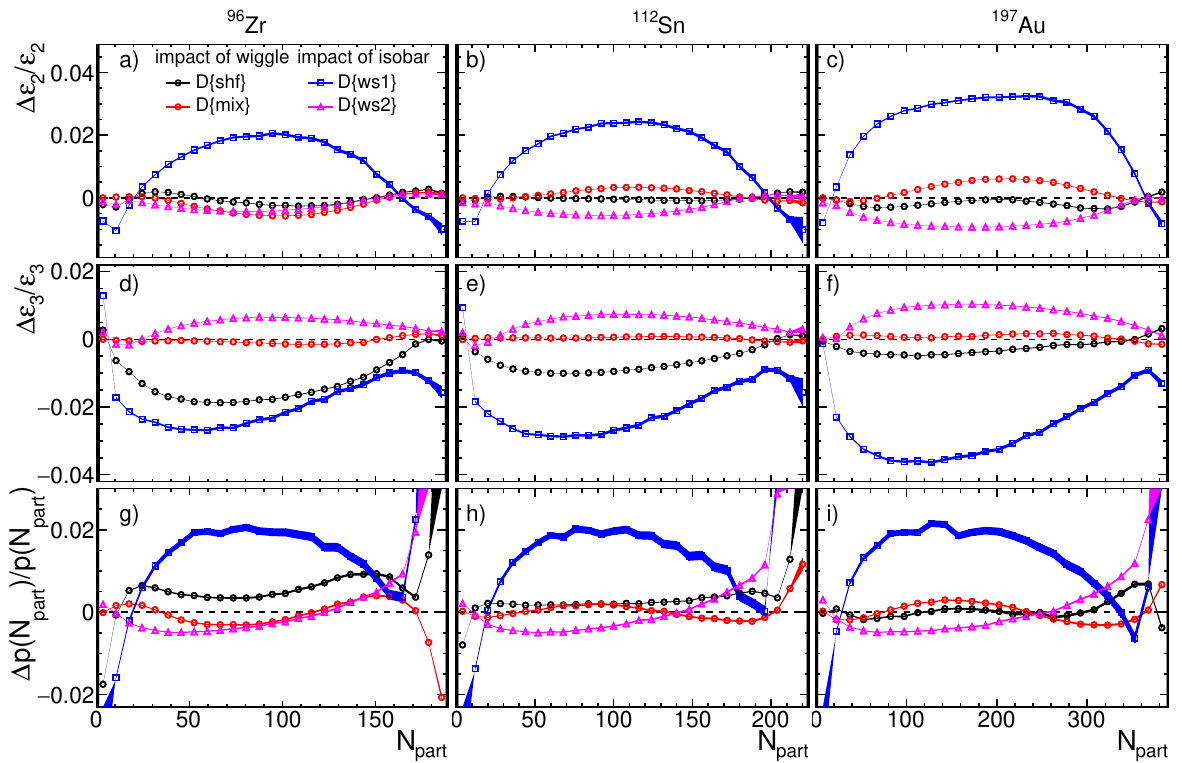}    
\caption{\label{fig:2} Glauber model results of $D_{\mathcal{O}}$ for $\mathcal{O}=\varepsilon_2$ (top row), $\varepsilon_3$ (middle row), and $p(\npart)$ (bottom row) for Zr+Zr (left column), Sn+Sn (middle column) and Au+Au (right column) collisions. Four variations are shown: $D_{\mathcal{O}}\{\mathrm{shf}\}$ (SHF vs. WS), $D_{\mathcal{O}}\{\mathrm{mix}\}$ (Mixed vs. WS), $D_{\mathcal{O}}\{\mathrm{ws1}\}$ (WS with $a$ decreased by 0.06~fm vs. without) and $D_{\mathcal{O}}\{\mathrm{ws2}\}$ (WS with $R_0$ increased by 0.07~fm vs. without), see text for explanation.}
\end{figure*}

{\bf Results} Figure~\ref{fig:2} shows the relative differences,  as functions of $\npart$ for the three observables. We first discuss the expected signal from isobar ratios. In $^{96}$A+$^{96}$A collisions, reducing $a$ by $0.06$~fm, represented by $D_{\mathcal{O}}\{\mathrm{ws1}\}$, leads to a 2\% enhancement in $\varepsilon_2$ in mid-central collisions, while increasing $R_0$ by $0.07$~fm, represented by $D_{\mathcal{O}}\{\mathrm{ws2}\}$, decreases $\varepsilon_2$ by about 0.4\%. These behaviors align with the expectation that a more compact source increases $\varepsilon_2$~\cite{Jia:2022qgl}. The variations in both cases are smaller in peripheral and central collisions. For larger $^{197}$A+$^{197}$A system, the impact is more significant: the same variations in $a$ and $R_0$ lead to a 3.3\% increase and a 0.9\% decrease in $\varepsilon_2$, respectively. The reason is that larger systems have smaller $\varepsilon_2$ values, amplifying the influence of WS parameter variations. A significant impact is observed for $\varepsilon_3$ (middle row of Fig.~\ref{fig:2}), with the relative change having a sign opposite to that for $\varepsilon_2$. These findings are qualitatively consistent with previous studies~\cite{Shou:2014eya,Xu:2017zcn}.

The impact on $p(\npart)$ can be understood as follows: reducing $a$ sharpens the overlap region's edge, increasing the probability for mid-central and ultra-central events, resulting in a broad bump in the ratio, similar to the STAR isobar data~\cite{STAR:2021mii}. The bump's magnitude is purely geometrical and nearly independent of the system size. In contrast, increasing $R_0$ slightly reduces $p(\npart)$ in mid-central collisions.

\begin{figure*}[htbp]
\includegraphics[width=0.9\linewidth]{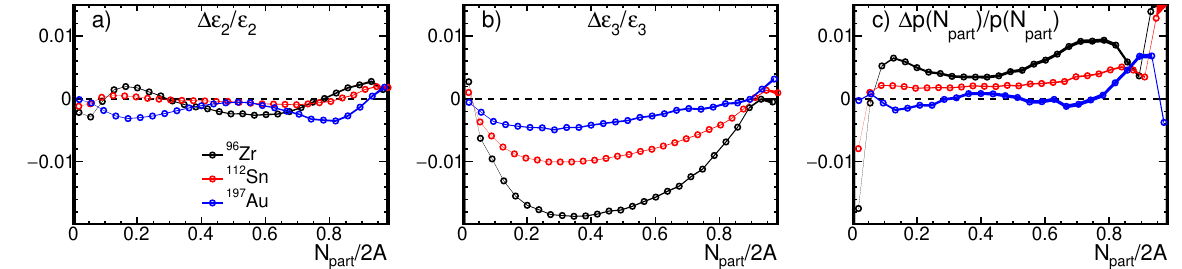}    
\caption{\label{fig:3} Relative difference $D_{\mathcal{O}}=\Delta\mathcal{O}/\mathcal{O}$ between SHF and WS with moment matching, where $\mathcal{O}$ is $\varepsilon_2$ (left), $\varepsilon_3$ (middle), and $p(\npart)$ (right) in Zr+Zr, Sn+Sn and Au+Au collisions, taken directly from Fig.~\ref{fig:2}. They are plotted as a function of $\npart/2A$, where A is the mass number.}
\end{figure*}
\begin{figure*}[htbp]
\includegraphics[width=0.9\linewidth]{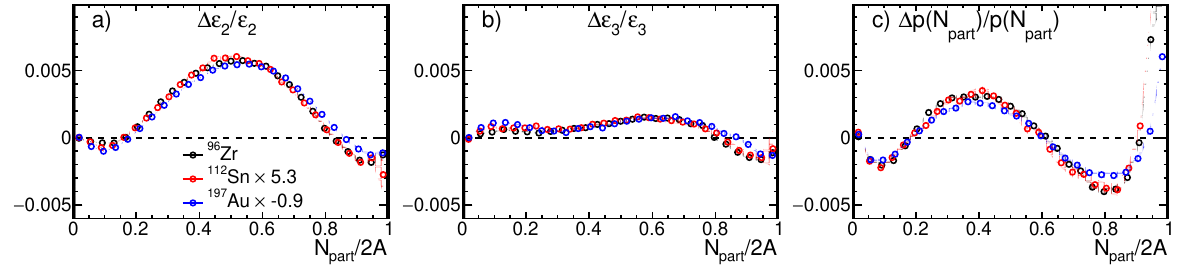}    
\caption{\label{fig:4} Relative difference $D_{\mathcal{O}}=\Delta\mathcal{O}/\mathcal{O}$ between $\rho_{\mathrm{mix}}$ and  $\rho_{\mathrm{WS}}$ (i.e differing only in the inner region), where $\mathcal{O}$ is $\varepsilon_2$ (left), $\varepsilon_3$ (middle), and $p(\npart)$ (right) in Zr+Zr, Sn+Sn, and Au+Au collisions, taken directly from Fig.~\ref{fig:2}. They are plotted as a function of $\npart/2A$, where A is the mass number.}
\end{figure*}

We then discuss the differences between SHF and WS, represented by $D_{\mathcal{O}}\{\mathrm{shf}\}$. They are represented by open black symbols in Fig.~\ref{fig:2}, which are also more conveniently compared across systems in Fig.~\ref{fig:3}. The impact on $\varepsilon_2$, which has an $r^2$ weight in its definition (Eq.~\eqref{eq:4b}), is minimal. However, $\varepsilon_3$ from SHF is about 2\% smaller than from WS in mid-central collisions, possibly due to the $r^3$ weight in $\varepsilon_3$ enhancing contributions from nucleons at the edge of the overlap region. The impact on $p(\npart)$ is small except in $^{96}$Zr, reaching about 0.4--0.6\% in mid-central collisions. This can be attributed to the larger outer bump around $R_0$ in Fig.~\ref{fig:1}b, more significant in $^{96}$Zr. The overall shape of $D_{\mathcal{O}}\{\mathrm{shf}\}$ shows similarities across different collision systems, particularly for $\varepsilon_3$.

Given the prevalent wiggly structures in $\rho_{\mathrm{shf}}/\rho_{\mathrm{ws}}$ (Fig.~\ref{fig:1}b), we investigate the impact of the inner region ($r<R_0-a$) and the surface region ($r>R_0-a$) of $\rho(r)$. For this, we examine $D_{\mathcal{O}}\{\mathrm{mix}\}$, which includes only the inner region's impact. These results, represented by open red symbols in Fig.~\ref{fig:2} and compared across systems in Fig.~\ref{fig:4}, show small magnitudes ($<0.5$\%) and are observable-dependent, but they exhibit the same $\npart$-dependent shapes across collision systems. The difference between $D_{\mathcal{O}}\{\mathrm{shf}\}$ and $D_{\mathcal{O}}\{\mathrm{mix}\}$ (Fig.~\ref{fig:2}) captures the surface region's contribution, which dominates the $\varepsilon_3$ differences between SHF and WS.

Finally, we consider whether the differences in $\varepsilon_2$, $\varepsilon_3$, or $p(\npart)$ between SHF and WS can be absorbed by adjusting the WS parameters $R_0$ and $a$ alone:
\small\begin{align}\nonumber
 D_{\mathcal{O}}\{\mathrm{shf}\}  &= a_1 D_{\mathcal{O}}\{\mathrm{ws1}\}+a_2 D_{\mathcal{O}}\{\mathrm{ws2}\}\\\label{eq:5}
 D_{\mathcal{O}}\{\mathrm{mix}\} &= b_1 D_{\mathcal{O}}\{\mathrm{ws1}\}+b_2 D_{\mathcal{O}}\{\mathrm{ws2}\}\;,
\end{align} \normalsize
where $a$ and $b$ are numerical coefficients common to all observables for a given collision species. This adjustment can only be done for one observable at a time. For example, the change in $\varepsilon_2$ in Fig.~\ref{fig:4}a can be reproduced by varying $a$, but this would induce a reduction in $\varepsilon_3$ and an enhancement in $p(\npart)$, which are not observed. Therefore, the influence of these fine structures on observables is real and can only be accurately studied via realistic DFT calculations.

Overall, the impact of DFT's wiggly structures on observables is smaller than (comparable to) typical variations of $a$ ($R_0$)  in isobar collisions, assuming they are similar to those between Zr+Zr and Ru+Ru. Including wiggly structures in SHF induces changes that are about four times smaller for $\varepsilon_2$ and twice smaller for $p(\npart)$ compared to a $0.06$~fm variation in $a$. However, the influences are comparable for $\varepsilon_3$. Thus, a WS fit of DFT is sufficient for studying the structure impact on $\varepsilon_2$ and $p(\npart)$ but not $\varepsilon_3$~\footnote{However, $\varepsilon_3$ is not a good estimator for studying the nuclear structure impact on the $v_3$ in peripheral and mid-central region (see bottom panels of Fig.~2 in Ref.~\cite{Jia:2021oyt} and also \cite{Nijs:2021kvn}). Therefore, it remains an open question whether considering WS parameters alone is adequate for $v_3$ in peripheral and mid-central regions.}.

{\bf Discussion and summary}.  We have demonstrated that the Woods-Saxon (WS) parameterization used to describe nucleon density distributions in heavy nuclei fails to capture the detailed radial wiggle structure in realistic Density Functional Theory (DFT) calculations. The differences between DFT and WS result in characteristic variations in observables for the initial state of heavy-ion collisions, such as $\varepsilon_2$, $\varepsilon_3$, and $p(\npart)$. These differences, on a few percent level, are centrality- and observable-dependent, with similar shapes across different nuclei but distinct behaviors among the three observables. These variations cannot be simultaneously absorbed by merely re-adjusting the WS parameters. This conclusion also applies to the final state observables $v_2$, $v_3$, and $p(\nch)$, thanks to the linear response relations $v_n\propto \varepsilon_n$ and $\nch \propto \npart$.

The magnitude of the changes induced by the differences between DFT and WS, on the order of less than a few percent, is much smaller than the typical experimental uncertainties for observables in a single collision system. For this purpose, WS parameterization is probably sufficient. However, since isobar ratios are extremely sensitive to nuclear structure differences, not considering DFT limits the precision with which nuclear structure can be constrained. In this case, more realistic DFT calculations should be used directly to gauge the systematic uncertainties.

Our work, together with earlier ones~\cite{Shou:2014eya,Xu:2021uar}, is a first step in quantifying whether the WS parameterization is sufficient in heavy-ion phenomenology. Future directions include considering the influence of nuclear deformation in DFT versus deformed WS parameterization~\cite{Ryssens:2023fkv}, as well as studying other bulk observables such as mean transverse momentum fluctuations. More importantly, simulations based on state-of-the-art hydrodynamic models are required to quantify the impact of realistic DFT versus WS on the initial condition and the extraction of QGP's transport properties. The uncertainties in nuclear structure input should be included in the Bayesian analysis for heavy-ion collisions along the line of Ref.~\cite{Giacalone:2023cet}.

We thank G. Giacalone for useful comments. Z. Y and J.J are supported by DOE under Grants No. DE-SC0024602. JX is supported by the Strategic Priority Research Program of the Chinese Academy of Sciences under Grant No. XDB34030000, the NSFC under Grants No. 12375125, and the Fundamental Research Funds for the Central Universities.

\section*{Appendix}
\begin{figure}[!h]
\includegraphics[width=0.8\linewidth]{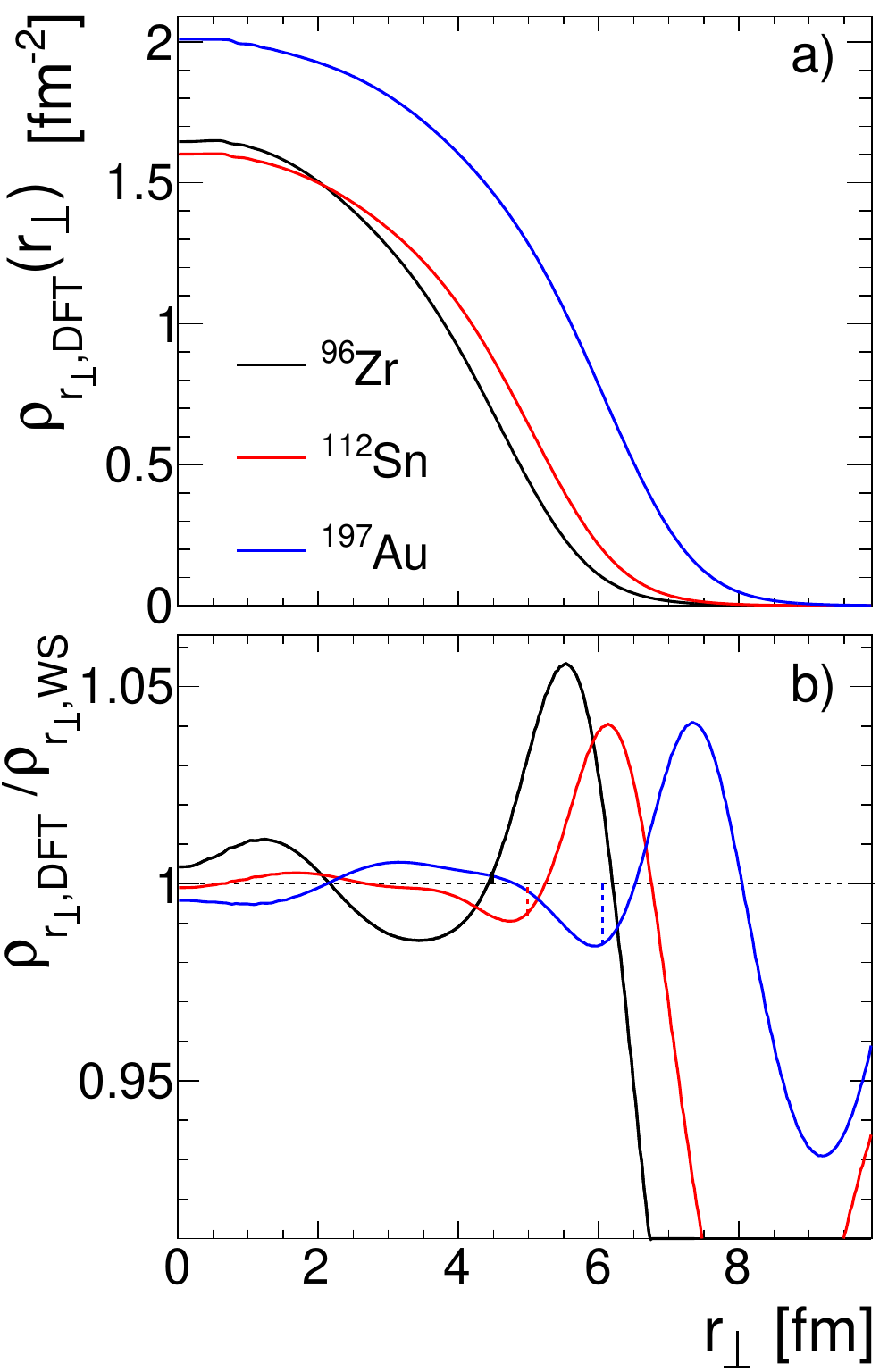}
\caption{\label{fig:a0} (a) SHF nucleon density distributions in the transverse plane for $^{96}$Zr, $^{112}$Sn, and $^{197}$Au nuclei. (b) Ratios of SHF and WS density distributions after moment matching and projection to the transverse plane. The vertical dashed lines indicate the location $r_{\perp}=R_0-a$.}
\end{figure}

Figure~\ref{fig:a0}a shows the nucleon radial density distribution in the transverse plane, $\rho_{\perp}(r_\perp)=\int \rho dzd\phi$ for SHF. This distribution is directly used to calculate $\varepsilon_n$ and $\npart$. Notably, in the interior region, the density of $^{96}$Zr remains higher than that of $^{112}$Sn, even after integration along the z-axis. The corresponding ratios between the SHF and WS distributions in the transverse plane are shown in Fig.~\ref{fig:a0}b. Compared to Fig.~\ref{fig:1}b, the wiggle structures at $r_\perp \lesssim R_0$ are reduced but not eliminated, while the structures at $r_\perp \gtrsim R_0$ remain more or less unchanged.

\bibliography{decay}{}
\end{document}